%% file: sbmf_entcs.tex
\documentclass{entcs}
\usepackage{entcsmacro}
\usepackage{amsmath,amsfonts}
\usepackage{epsfig}
\usepackage{url}

\def\lastname{Tafliovich}
\begin{document}
\begin{frontmatter}
  \title{Programming Telepathy: Implementing Quantum Non-locality
    Games} \author{Anya Tafliovich\thanksref{ALL}\thanksref{myemail}}
  \address{Computer Science\\ University of Toronto\\
    Toronto, Canada} \author{Eric C. R. Hehner\thanksref{coemail}}
  \address{Computer Science\\University of Toronto\\
    Toronto, Canada} \thanks[ALL]{This work is in part supported by
    NSERC} \thanks[myemail]{Email: \href{mailto:anya@cs.toronto.edu}
    {\texttt{\normalshape anya@cs.toronto.edu}}}
  \thanks[coemail]{Email: \href{mailto:hehner@cs.toronto.edu}
    {\texttt{\normalshape hehner@cs.toronto.edu}}}
\begin{abstract} 
  Quantum pseudo-telepathy is an intriguing phenomenon which results
  from the application of quantum information theory to communication
  complexity. To demonstrate this phenomenon researchers in the field
  of quantum communication complexity devised a number of quantum
  non-locality games. The setting of these games is as follows: the
  players are separated so that no communication between them is
  possible and are given a certain computational task. When the
  players have access to a quantum resource called entanglement, they
  can accomplish the task: something that is impossible in a classical
  setting. To an observer who is unfamiliar with the laws of quantum
  mechanics it seems that the players employ some sort of telepathy;
  that is, they somehow exchange information without sharing a
  communication channel.

  This paper provides a formal framework for specifying, implementing,
  and analysing quantum non-locality games. 
\end{abstract}
\end{frontmatter}

\include{defs}

%%%%%%%%%%%%%%%%%%%%%%%%%%%%%%%%%%%%%%%%%%%%%%%%%%%%%%%%%%%%%%%%%%%%%%
% Introduction
%%%%%%%%%%%%%%%%%%%%%%%%%%%%%%%%%%%%%%%%%%%%%%%%%%%%%%%%%%%%%%%%%%%%%%

\section{Introduction}
\label{sec:introduction}

The work develops a formal framework for specifying, implementing,
and analysing quantum pseudo-telepathy: an intriguing phenomenon which
manifests itself when quantum information theory is applied to
communication complexity.  To demonstrate this phenomenon researchers
in the field of quantum communication complexity devised a number of
quantum non-locality games. The setting of these games is as follows:
the players are separated so that no communication between them is
possible and are given a certain computational task. When the players
have access to a quantum resource called entanglement, they can
accomplish the task: something that is impossible in a classical
setting. To an observer who is unfamiliar with the laws of quantum
mechanics it seems that the players employ some sort of telepathy;
that is, they somehow exchange information without sharing a
communication channel.

Quantum pseudo-telepathy, and quantum non-locality in general, are
perhaps the most non-classical and the least understood aspects of
quantum information processing. Every effort is made to gain
information about the power of these phenomena. Quantum non-locality
games in particular have been extensively used to prove separations
between quantum and classical communication complexity. The need for a
good framework for formal analysis of quantum non-locality is evident.

We look at quantum non-locality in the context of formal methods of
program development, or programming methodology. This is the field of
computer science concerned with applications of mathematics and logic
to software engineering tasks. In particular, the formal methods
provide tools to formally express specifications, prove correctness of
implementations, and reason about various properties of specifications
(e.g. implementability) and implementations (e.g. time and space
complexity).

In this work the analysis of quantum non-locality is based on quantum
predicative programming (\cite{tafliovich06,tafliovich04}), a recent
generalisation of the well-established predicative programming
(\cite{hehner:book,hehner04,hehner09}).  It supports the style of
program development in which each programming step is proved correct
as it is made.  We inherit the advantages of the theory, such as its
generality, simple treatment of recursive programs, and time and space
complexity. The theory of quantum programming provides tools to write
both classical and quantum specifications, develop quantum programs
that implement these specifications, and reason about their
comparative time and space complexity all in the same framework.

Presenting new non-locality paradigms or new pseudo-telepathy games is
not the subject of this work. Our goal is developing a formal
framework that encompasses all aspects of quantum computation and
information.  Formal analysis of quantum algorithms, including their
time complexity, is presented in~\cite{tafliovich06}. Analysis of
quantum communication appears in~\cite{tafliovich09}. This paper
focuses on formal analysis of non-locality paradigms; we choose known
pseudo-telepathy games as illustrative examples of our formalism.

The rest of this work is organised as follows.  Section~\ref{sec:qpp}
is a brief introduction to quantum predicative programming.  The
contribution of this work is Section~\ref{sec:nonlocality} which
introduces a formal framework for specifying, implementing, and
analysing quantum pseudo-telepathy and presents several examples of
implementing and analysing non-locality games.
Section~\ref{sec:conclusion} states conclusions and outlines
directions for future research. A brief introduction to quantum
computing is included in the Appendix.

\subsection{Our contribution and related work}
\label{sec:relwork}

This work attempts to bring together two areas of active research: the
study of quantum non-locality and applications of formal methods to
quantum information and computation. Currently, the two worlds rarely
meet.
 
Quantum non-locality has been studied extensively first by physicists
and lately by researchers in the fields of quantum information and
quantum communication complexity.  Since the work of Bell in 1964
(\cite{bell64}), researchers have been trying to provide an intuitive
explanation of the genuinely non-classical behaviour produced by
quantum mechanics. Today, quantum pseudo-telepathy games are
considered one of the best and easiest to understand examples of these
non-classical phenomena
(e.g.~\cite{galliard02,brassard04,brassard05,brassard06}).

Formal approaches to quantum programming include the language
qGCL~\cite{sanders00,zuliani04,zuliani05}, process algebraic
approaches developed in~\cite{adao,lalire04,jorrand04}, tools
developed in the field of category theory
by~\cite{abramsky04,abramsky04:csqp,abramsky06,coecke04:le,selinger04},
functional languages
of~\cite{arrighi04,arrighi05,altenkirch05:fqpl,valiron04,vanTonder04},
as well as work of~\cite{dHondt06:wp,dHondt05},~\cite{danos05},
and~\cite{gay05}. A detailed discussion of the work related to quantum
predicative programming is presented in~\cite{tafliovich06}. Some
researchers address the subject of formalising quantum non-locality
more directly than others (e.g.~\cite{zuliani04}). To the best of our
knowledge, formal approaches to reasoning about quantum
pseudo-telepathy games have not been considered.

%%%%%%%%%%%%%%%%%% end Introduction %%%%%%%%%%%%%%%%%%%%%%%%%%%%%%%%%%

%%%%%%%%%%%%%%%%%%%%%%%%%%%%%%%%%%%%%%%%%%%%%%%%%%%%%%%%%%%%%%%%%%%%%%
% Quantum Predicative Programming
%%%%%%%%%%%%%%%%%%%%%%%%%%%%%%%%%%%%%%%%%%%%%%%%%%%%%%%%%%%%%%%%%%%%%%

\section {Quantum Predicative Programming}
\label{sec:qpp}

This section introduces the programming theory of our choice ---
quantum predicative programming. We briefly introduce parts of the
theory necessary for understanding Section~\ref{sec:nonlocality} of
this work. For a course in predicative programming the reader is
referred to~\cite{hehner:book}. An introduction to probabilistic
predicative programming can be found in~\cite{hehner04,hehner09}.
Quantum predicative programming is developed
in~\cite{tafliovich06,tafliovich04} and is extended with reasoning
about quantum communication in~\cite{tafliovich09}.

\subsubsection{Predicative programming}
\label{sec:pp}

In predicative programming a specification is a boolean expression. The
variables in a specification represent the quantities of interest,
such as prestate (inputs), poststate (outputs), and computation time
and space. We use primed variables to describe outputs and unprimed
variables to describe inputs.  For example, specification $x' = x + 1$
in one integer variable $x$ states that the final value of $x$ is its
initial value plus $1$.  A computation \emph{satisfies} a
specification if, given a prestate, it produces a poststate, such that
the pair makes the specification true. A specification is
\emph{implementable} if for each input state there is at least one
output state that satisfies the specification.

We use standard logical notation for writing specifications: $\wedge$
(conjunction), $\vee$ (disjunction), $\Rightarrow$ (logical
implication), $=$ (equality, boolean equivalence), $\neq$
(non-equality, non-equivalence), and \textbf{if then else}. The larger
operators $\eqv$, $\Longrightarrow$, $\leqv$, and $\geqv$ are the same
as $=$, $\Rightarrow$, $\leq$, and $\geq$, but with lower precedence.
We use standard mathematical notation, such as $ + \, - \, \times \, /
\, mod $.  We use lowercase letters for variables of interest and
uppercase letters for specifications.

In addition to the above, we use the following notations: $\sigma$
(prestate), $\sigma'$ (poststate), $ok$ ($\sigma'=\sigma$), and $x:=e$
defined by $x'=e \wedge y'=y \wedge \hdots$. The notation $ok$
specifies that the values of all variables are unchanged. In the
assignment $x:=e$, $x$ is a state variable (unprimed) and $e$ is an
expression (in unprimed variables) in the domain of $x$.

If $R$ and $S$ are specifications in variables $x, y, \hdots \;$, then
the \emph{sequential composition} of $R$ and $S$ is defined by
$$R;S \eqv \exists x'', y'', \hdots \cdot R'' \wedge S''$$
where $R''$ is obtained from $R$ by substituting all occurrences of
primed variables $x', y', \hdots$ with double-primed variables $x'',
y'', \hdots \;$, and $S''$ is obtained from $S$ by substituting all
occurrences of unprimed variables $x, y, \hdots$ with double-primed
variables $x'', y'', \hdots \;$.

Various laws can be proved about sequential composition. One of the
most important ones is the substitution law, which states that for any
expression $e$ of the prestate, state variable $x$, and specification
$P$,

$$x:=e; P \eqv (\text{for } x \text{ substitute } e \text{ in } P)$$

Specification $S$ \emph{is refined by} specification $P$ if and only
if $S$ is satisfied whenever $P$ is satisfied, that is $\forall
\sigma, \sigma' \cdot S \Leftarrow P$. Given a specification, we are
allowed to implement an equivalent specification or a stronger one.

A \emph{program} is an implemented specification. A good basis for
classical (non-quantum) programming is provided by: $ok$, assignment,
\textbf{if then else}, sequential composition, booleans, numbers,
bunches, and functions.

Given a specification $S$, we proceed as follows. If $S$ is a program,
there is no work to be done. If it is not, we build a program $P$,
such that $P$ refines $S$, i.e. $S \Leftarrow P$. The refinement can
proceed in steps: $S \Leftarrow \hdots \Leftarrow R \Leftarrow Q
\Leftarrow P$.

In $S \Leftarrow P$ it is possible for $S$ to appear in $P$. No
additional rules are required to prove the refinement. For example, it
is trivial to prove that
\begin{equation*}
  x \geq 0 \Rightarrow x'=0 
  \lar
  \textbf{if } x=0
  \textbf{ then } ok 
  \textbf{ else } (x:=x-1\seq x \geq 0 \Rightarrow x'=0 )
\end{equation*}

The specification says that if the initial value of $x$ is
non-negative, its final value must be $0$. The solution is: if the
value of $x$ is zero, do nothing, otherwise decrement $x$ and repeat.

\subsubsection{Probabilistic predicative programming}
\label{sec:ppp}

A \emph{probability} is a real number between $0$ and $1$, inclusive.
A \emph{distribution} is an expression whose value is a probability
and whose sum over all values of variables is $1$. Given a
distribution of several variables, we can sum out some of the
variables to obtain a distribution of the rest of the variables.

To generalise boolean specifications to probabilistic specifications,
we use $1$ and $0$ both as numbers and as boolean $\mathit{true}$ and
$\mathit{false}$, respectively.\footnote{Readers familiar with $\top$
  and $\bot$ notation can notice that we take the liberty to equate
  $\top=1$ and $\bot=0$.} If $S$ is an implementable deterministic
specification and $p$ is a distribution of the initial state $x, y,
...$, then the distribution of the final state is
\begin{equation*}
  \sum {x, y, ...} \cdot S \times p
\end{equation*}

If $R$ and $S$ are specifications in variables $x, y, \hdots \;$, then
the \emph{sequential composition} of $R$ and $S$ is defined by
\begin{equation*}
  R;S \eqv \sum {x'', y'', \hdots} \cdot R'' \times S''
\end{equation*}
where $R''$ is obtained from $R$ by substituting all occurrences of primed
variables $x', y', \hdots$ with double-primed variables $x'', y'',
\hdots \;$, and $S''$ is obtained from $S$ by substituting all
occurrences of unprimed variables $x, y, \hdots$ with double-primed
variables $x'', y'', \hdots \;$.

If $p$ is a probability and $R$ and $S$ are distributions, then
\begin{equation*}
  \textbf{if } p \textbf{ then } R \textbf{ else } S \eqv
  p \times R + (1-p)\times S
\end{equation*}

Various laws can be proved about sequential composition. One of the
most important ones, the substitution law, introduced earlier, applies
to probabilistic specifications as well.

We use assignment, sequential composition, and if-then-else to reason
about probability distributions that result from (probabilistically)
changing the state variables. To reason about probability
distributions that result from learning some new information, with no
change to the state variables, we use the learn operator, introduced
in~\cite{hehner09}. If $P$ is the original probability distribution
and $b$ is a boolean expression that describes the information we
learn, then the resulting probability distribution is defined by
$$ P \lrn b \eqv (P \times b') / (P \seq b)$$

If $P$ is a non-negative expression (not necessarily a probability
distribution), then $P \lrn 1$ is the normalisation of $P$.

If $P$ and $Q$ are distributions, then $P \leq Q$ is the
generalisation of refinement $P \Rightarrow Q$ to the probabilistic
case.

\subsubsection{Quantum Predicative Programming}
\label{sec:qpp_sub}

Let $\mathbb{C}$ be the set of all complex numbers with the absolute
value operator $|\cdot|$ and the complex conjugate operator $^*$. Then
a state of an $n$-qubit system is a function $\psi : 0,..2^n
\rightarrow \mathbb{C}$, such that $\sum {x:0,..2^n} \cdot |\psi x|^2
\eqv 1$.

If $\psi$ and $\phi$ are two states of an $n$-qubit system, then their
\emph{inner product}, denoted by $\braket{\psi}{\phi}$, is
defined by\footnote{We should point out that this kind of function
  operations is referred to as \emph{lifting}.}:
\begin{equation*}
  \braket{\psi}{\phi} = \sum {x:0,..2^n} \cdot (\psi x)^* \times (\phi x)
\end{equation*}

A \emph{basis} of an $n$-qubit system is a collection of $2^n$ quantum
states $b_{0,..2^n}$, such that $\forall i,j:0,..2^n \cdot
\braket{b_i}{b_j} = (i=j)$. We adopt the following Dirac-like notation
for the computational basis: if $x$ is from the domain $0,..2^n$, then
$\textbf{x}$ denotes the corresponding $n$-bit binary encoding of $x$
and $\ketb{x}: 0,..2^n \rightarrow \mathbb{C}$ is the following
quantum state:
\begin{equation*}
  \ketb{x} = \lambda i:0,..2^n \cdot (i=x)
\end{equation*}

If $\psi$ is a state of an $m$-qubit system and $\phi$ is a state of
an $n$-qubit system, then $\psi \tensor \phi$, the tensor product of
$\psi$ and $\phi$, is the following state of a composite $m+n$-qubit
system:
\begin{equation*}
  \psi \tensor \phi = \lambda i:0,..2^{m+n} \cdot 
                      \psi (i\:div\:2^n) \times \phi (i\:mod\:2^n)
\end{equation*}

We write $\psi^{\tensor n}$ to mean $\psi$ \emph{tensored with itself
  $n$ times}.

An operation defined on an $n$-qubit quantum system is a higher-order
function, whose domain and range are maps from $0,..2^n$ to the
complex numbers. An \emph{identity} operation on a state of an
$n$-qubit system is defined by
\begin{equation*}
  I^n = \lambda \psi : 0,..2^n \rightarrow \mathbb{C} \cdot \psi
\end{equation*}

For a linear operation $A$, the \emph{adjoint} of $A$, written
$A^\dagger$, is the (unique) operation, such that for any two states
$\psi$ and $\phi$, $\braket{\psi}{A \phi} = \braket{A^{\dagger}
  \psi}{\phi}$.

The \emph{unitary transformations} that describe the evolution of an
$n$-qubit quantum system are operations $U$ defined on the system,
such that $U^{\dagger} U = I^n$.

In this setting, the \emph{tensor product} of operators is defined in
the usual way. If $\psi$ is a state of an $m$-qubit system, $\phi$ is
a state of an $n$-qubit system, and $U$ and $V$ are operations defined
on $m$ and $n$-qubit systems, respectively, then the tensor product of
$U$ and $V$ is defined on an $m+n$ qubit system by $(U \tensor V)
(\psi \tensor \phi) = (U \psi) \tensor (V \phi)$.

Just as with tensor products of states, we write $U^{\tensor n}$ to
mean \emph{operation $U$ tensored with itself $n$ times}.

Suppose we have a system of $n$ qubits in state $\psi$ and we measure
it. Suppose also that we have a variable $r$ from the domain
$0,..2^n$, which we use to record the result of the measurement, and
variables $x,y, \hdots$, which are not affected by the measurement.
Then the measurement corresponds to a probabilistic specification that
gives the probability distribution of $\psi'$ and $r'$ (these depend
on $\psi$ and on the type of measurement) and states that the
variables $x,y,\hdots$ are unchanged.

For a general quantum measurement described by a collection $M =
M_{0,..2^n}$ of measurement operators, which satisfy the completeness
equation (see Appendix~\ref{sec:qc}), the specification is
$\textbf{measure}_M \, \psi \, r$, where
\begin{equation*}
  \textbf{measure}_M \, \psi \, r \eqv
    \braket{\psi}{M_{r'}^\dagger M_{r'}\psi} \times
    \left(
     \psi'=\frac {M_{r'}\psi}{\sqrt{\braket{\psi}
                 {M_{r'}^\dagger M_{r'}\psi}}}
    \right) \times (\sigma'=\sigma)
\end{equation*}
where $\sigma'=\sigma$ is an abbreviation of $(x'=x) \times (y'=y)
\times \hdots$ and means ``all other variables are unchanged''.

The simplest and the most commonly used measurement in the
computational basis is:
\begin{equation*}
  \textbf{measure } \psi \, r \eqv
    | \psi r' |^2  \times (\psi'= |\textbf{r}'\rangle) \times (\sigma'=\sigma)
\end{equation*}

In this case the distribution of $r'$ is $|\psi r'|^2$ and the
distribution of the quantum state is:
\begin{equation*}
  \sum {r'} \cdot |\psi r'|^2 \times (\psi' = |\textbf{r}'\rangle)
\end{equation*}
which is precisely the mixed quantum state that results from the
measurement. 

In order to develop quantum programs we need to add to our list of
implemented things. We add variables of type quantum state as above
and we allow the following three kinds of operations on these
variables.  If $\psi$ is a state of an $n$-qubit quantum system, $r$
is a natural variable, and $M$ is a collection of measurement
operators that satisfy the completeness equation, then:
\begin{enumerate}
\item $\psi:=\ket{0}^{\tensor n}$ is a program
\item $\psi:=U\psi$, where $U$ is a unitary transformation on an
  $n$-qubit system, is a program
\item $\textbf{measure}_M \, \psi \, r$ is a program
\end{enumerate}
The special cases of measurements are therefore also allowed.

The \emph{Hadamard} transform, widely used in quantum algorithms, is
defined on a $1$-qubit system and in our setting is a higher-order
function on $0,1 \rightarrow \mathbb{C}$:
\begin{equation*}
  H = \lambda \psi:0,1 \rightarrow \mathbb{C} \cdot \lambda i:0,1 \cdot
  (\psi 0 + (-1)^i \times  \psi 1)/\sqrt2
\end{equation*}

The operation $H^{\tensor n}$ on an $n$-qubit system applies $H$ to
every qubit of the system. Its action on the zero state of an
$n$-qubit system is:
\begin{equation*}
  H^{\tensor n} \ket{0}^{\tensor n} =
     \sum {x:0,..2^n} \cdot \ketb{x} / \sqrt{2^n}
\end{equation*}

On a basis state $\ketb{x}$, the action of $H^{\tensor n}$ is:
\begin{equation*}
  \label{eq:hadamard_on_x}
  H^{\tensor n} \ketb{x} =
    \sum {y:0,..2^n} \cdot 
      (-1)^{\textbf{x} \cdot \textbf{y}} \times \ketb{y} / \sqrt{2^n}
\end{equation*}
where $\textbf{x} \cdot \textbf{y}$ is the inner product of 
$\textbf{x}$ and $\textbf{y}$ modulo 2.

%%%%%%%%%%%%%%%%%%%%%%%%     end QPP   %%%%%%%%%%%%%%%%%%%%%%%%%%%%%%%

%%%%%%%%%%%%%%%%%%%%%%%%%%%%%%%%%%%%%%%%%%%%%%%%%%%%%%%%%%%%%%%%%%%%%%
% Quantum Non-locality
%%%%%%%%%%%%%%%%%%%%%%%%%%%%%%%%%%%%%%%%%%%%%%%%%%%%%%%%%%%%%%%%%%%%%%
\section{Quantum Non-locality}
\label{sec:nonlocality}

In predicative programming, to reason about distributed computation we
(disjointly) partition the variables between the processes involved in
a computation. Parallel composition is then simply boolean conjunction.
For example, consider two processes $P$ and $Q$. $P$ owns integer
variables $x$ and $y$ and $Q$ owns an integer variable $z$. Suppose
$ P \eqv x:=x+1; y:=x \text{ and } Q \eqv z:=-z $.
Parallel composition of $P$ with $Q$ is then simply
\begin{align*}
  & P || Q \eqv P \land Q \\
  \eqv
  & (x \get x+1 \seq y \get x) \land (z \get -z)\\
  \eqv
  & x'=x+1 \land y'=x+1 \land z'=-z
\end{align*}

In quantum predicative programming, one needs to reason about
distributed quantum systems. Recall that if $\psi$ is a state of an
$m$-qubit system and $\phi$ is a state of an $n$-qubit system, then
$\psi \tensor \phi$, the tensor product of $\psi$ and $\phi$, is the
state of a composite $m+n$-qubit system. On the other hand, given a
composite $m+n$-qubit system, it is not always possible to describe it
in terms of the tensor product of the component $m$- and $n$-qubit
systems. Such a composed system is \emph{entangled}. Entanglement is
one of the most non-classical, most poorly understood, and most
interesting quantum phenomena. An entangled system is in some sense
both distributed and shared. It is distributed in the sense that each
party can apply operations and measurements to only its qubits. It is
shared in the sense that the actions of one party affect the outcome of
the actions of another party. Simple partitioning of qubits is
therefore insufficient to reason about distributed quantum
computation. 

The formalism we introduce fully reflects the physical properties of a
distributed quantum system. We start by partitioning the qubits
between the parties involved. For example, consider two parties $P$
and $Q$. $P$ owns the first qubit of the composite entangled quantum
system $\psi = \ket{00}/\sqrt{2} + \ket{11}/\sqrt{2}$ and $Q$ owns the
second qubit. A specification is a program only if each party computes
with its own qubits. In our example, 
$$P \eqv \psi_0 := H \phi_0; \; \textbf{measure } \psi_0\; p
  \quad\text{and}\quad Q \eqv \textbf{measure } \psi_1\; q$$
are programs, if $p$ and $q$ are integer variables owned by $P$ and
$Q$, respectively. The parties $P$ and $Q$ can access only their own
qubits: they could in theory be light years apart.

We define parallel composition of $P$ and $Q$ which share an $n+m$
quantum system in state $\psi$ with the first $n$ qubits belonging to
$P$ and the other $m$ qubits belonging to $Q$ as follows. If
$$ P \eqv \psi_{0..n} := U_P \psi_{0..n} 
   \quad\text{and}\quad
   Q \eqv \psi_{n..n+m} := U_Q \psi_{n..n+m}$$
where $U_P$ is a unitary operation on an $n$-qubit system and $U_Q$ is
a unitary operation on an $m$-qubit system, then
$$ P\; ||_\psi\; Q \eqv \psi := (U_P \tensor U_Q) \psi$$

Performing $ok$ is equivalent to performing the identity unitary
operation, and therefore if
$$ P \eqv \psi_{0,..n} := U_P \psi_{0,..n} 
   \quad\text{and}\quad
   Q \eqv ok$$
then
$$ P\; ||_\psi\; Q \eqv \psi := (U_P \tensor I^{\tensor m}) \psi$$

Similarly, if
$$ P \eqv \textbf{measure}_{M_P}\; \psi_{0..n} \;p 
   \quad\text{and}\quad
   Q \eqv \textbf{measure}_{M_Q}\; \psi_{n..n+m} \;q$$
where $M_P$ and $M_Q$ are a collection of proper measurement operators
for $n$- and $m$-qubit systems, respectively, then
$$P\; ||_\psi\; Q \eqv \textbf{measure}_{M_P \tensor M_Q} \psi \; pq$$
where $pq$ is the number that corresponds to the binary string
\textbf{pq}.

In our example,
\begin{alignat*}{2}
  & \psi := \ket{00}/\sqrt{2} + \ket{11}/\sqrt{2}; \;
     P \;||_\psi Q 
   &\text{expand, substitute}\br
\eqv
  & \psi := \ket{00}/\sqrt{2} + \ket{11}/\sqrt{2}; &\\
   & \textbf{measure } (H \psi_0)\; p 
     \;||_\psi\;
    \textbf{measure } \psi_1\; q
   &\text{compose on }\psi\br
\eqv
  & \psi := \ket{00}/\sqrt{2} + \ket{11}/\sqrt{2}; \;
    \textbf{measure } (H \tensor I) \psi \; pq
   &\text{substitute}\br
\eqv
  & \textbf{measure } (H \tensor I) 
      (\ket{00}/\sqrt{2} + \ket{11}/\sqrt{2}) \; pq
   &\text{apply }H \tensor I\br
\eqv
  & \textbf{measure }  
      (\ket{00} + \ket{01} + \ket{10} - \ket{11})/2 \; pq
   &\text{measure}\br
\eqv
  & \left|(\ket{00} + \ket{01} + \ket{10} - \ket{11})/2 \; pq \right|^2
    \times
    (\psi' = |\textbf{p}'\textbf{q}'\rangle)
   &\text{application}\br
\eqv
  & (\psi'= |\textbf{p}'\textbf{q}'\rangle)/4
\end{alignat*}

\section{Pseudo-telepathy games}

We formalise pseudo-telepathy games with $n$ players as follows. For
each player $i$, $0 \leq i < n$, we have a domain $D_i$ from which the
inputs to player $i$ are provided and a range $R_i$ of player $i$'s
possible output results. In addition we may have a promise $P$: a
condition on the inputs to the players. If no promise is given, we set
$P$ to $1$. The winning condition $W$ can involve inputs as well as
outputs for each player. The strategy $S$ is a program, i.e. an
implemented specification. The strategy $S$ is winning if, assuming
the promise, the strategy yields a distribution that corresponds to
the winning condition:
\begin{equation*}
  S \lrn P \leq W \lrn 1
\end{equation*}

\subsection{Deutsch-Jozsa game}
\label{sec:deutschjoszagame}

The Deutsch-Jozsa pseudo telepathy game~\cite{brassard99,brassard05}
is based on a well-known Deutsch-Jozsa algorithm~\cite{jozsa91}. The
setting of the game is as follows. Alice and Bob are separated several
light years apart and are each presented with a $2^k$-bit string. They
are promised that either the strings are identical or they differ by
exactly half of the bits.  To win the game the players must each
output a $k$-bit string, and these strings should be identical if and
only if their input strings were identical.

We formalise the game as follows. We partition the space into the
world of Alice (variables subscripted $A$) and the world of Bob
(variables subscripted $B$). Then we have the following formalisation
of the game:
\begin{align*}
  & D_A = D_B = \{0,1\}^{2^k} 
  &&\vspace{-5mm}\text{the domain of inputs}\\
  &R_A = R_B =\{0,1\}^k
  &&\vspace{-5mm}\text{the range of outputs}\\
  &P = P_0 \lor P_1
  &&\vspace{-5mm}\text{the promise on the inputs, where:}\\
  &P_0 \eqv x_A = x_B \\
  &\hphantom{P_0} \eqv (\sum i: 0,..2^k \cdot (x_A)_i = (x_B)_i) = 2^k
  &&\vspace{-5mm}\text{the inputs are identical}\\
  &P_1 \eqv (\sum i: 0,..2^k \cdot (x_A)_i =(x_B)_i) = 2^{k-1}
  &&\vspace{-5mm}\text{the inputs differ by half of the bits}\\
  &W \eqv P_0 \land (y_A'=y_B') \lor P_1 \land (y_A' \neq y_B')
  &&\vspace{-5mm}\text{the winning condition}
\end{align*}

We demonstrate the quantum solution by implementing the following
specification $S$:
\begin{align*}
  &S \eqv \psi \get  \sum z:0,..2^k \cdot \ketb{zz} / \sqrt{2^k}\seq  
          (S_A \;||_\psi\;  S_B) \quad\text{where}\\ 
  &S_i \eqv \psi_i \get  U_i^{\tensor k} \psi_i\seq
            \psi_i \get  H^{\tensor k}\psi_i\seq 
            \textbf{measure } \psi_i \;  y_i
            \quad\text{for unitary}\\ 
  &U_i \ketb{z} = (-1)^{(x_i)_z}\sqz{\times}\ketb{z}\qquad\text{where } i:A,B
\end{align*}

Implementing the initial assignment:
\begin{align*}
  &\psi \get  \sum z:0,..2^k \cdot \ketb{zz} / \sqrt{2^k}
  &\text{}\\
\eqv
  &\psi \get  \ket{0}^{\tensor 2k} \seq 
   \psi_{0,..k} \get H^{\tensor k} \psi_{0,..k} \seq
   \psi \get CNOT^{\tensor k}_d \psi 
\end{align*}

We begin by analysing the distribution that results from executing the
solution program (we omit domains of $u,v,z$ for clarity and sum out
the final quantum state, since it does not appear in the winning
condition)
\begin{alignat*}{2}
  &\sum \psi' \cdot S &\\
\eqv 
 &\sum \psi' \cdot 
  \psi \get \sum z \cdot \ketb{zz} / \sqrt{2^k}\seq 
   (S_A  \;||_\psi\;  S_B)
   &\text{expand } S_A, S_B\br
\eqv
 &\sum \psi' \cdot 
 \begin{aligned}[t]
   &\psi \get \sum z \cdot \ketb{zz} / \sqrt{2^k}\seq\\
   &((\psi_A \get  U_A^{\tensor k} \psi_A\seq
   \psi_A \get  H^{\tensor k}\psi_A\seq 
   \textbf{measure } \psi_A \;  y_A )
   \;||_\psi\; \\
   &\hphantom{(}(\psi_B \get  U_B^{\tensor k} \psi_B\seq
   \psi_B \get  H^{\tensor k}\psi_B\seq 
   \textbf{measure } \psi_B \;  y_B ))
 \end{aligned}
 &\begin{aligned}[t]
   &\\ &\\ &\text{substitute}
 \end{aligned}\br
\eqv
 &\sum \psi' \cdot 
 \begin{aligned}[t]
   &\psi \get \sum z \cdot \ketb{zz} / \sqrt{2^k}\seq\\ 
   &(\textbf{measure } H^{\tensor k} (U_A^{\tensor k} \psi_A) \;  y_A 
     \;||_\psi\;\\ 
   & \hphantom{(}\textbf{measure } H^{\tensor k} (U_B^{\tensor k} \psi_B) \;  y_B)
 \end{aligned}
 &\begin{aligned}[t]
   &&\\
   &&\text{composition}\\
   &&\quad\text{on $\psi$}
 \end{aligned}\br
\eqv
 &\sum \psi' \cdot 
 \begin{aligned}[t]
   &  \psi \get  \sum z \cdot \ketb{zz} / \sqrt{2^k}\seq \\
   & \textbf{measure } H^{\tensor 2k} 
   ((U_A^{\tensor k} \tensor U_B^{\tensor k}) \psi) \;  y_Ay_B
 \end{aligned}
 &\begin{aligned}[t]
   &&\text{substitute and}\\
   &&\text{measure}
 \end{aligned}\br
\eqv
  & \left|H^{\tensor 2k} \left((U_A^{\tensor k} \tensor U_B^{\tensor k}) 
    \left(\sum z \cdot \ketb{zz} / \sqrt{2^k}\right)\right) \;
      (y_Ay_B)'\right|^2 \;\times\\
    & (x_A'=x_A) \times (x_B'=x_B)
   &\text{linearity}\br
\eqv
  & \left|H^{\tensor 2k}  
      \left(\sum z \cdot 
             (U_A^{\tensor k}\ketb{z} \tensor
              U_B^{\tensor k}\ketb{z})\right) / \sqrt{2^k} \;
      (y_Ay_B)'\right|^2\times\\
    & (x_A'=x_A) \times (x_B'=x_B)
   &\text{apply }U_i\br
\eqv
  & \left|H^{\tensor 2k}  
      \left(\sum z \cdot 
            (-1)^{(x_A)_z}\sqz{\times}\ketb{z} \tensor
            (-1)^{(x_B)_z}\sqz{\times}\ketb{z} )\right) / \sqrt{2^k} \;
      (y_Ay_B)'\right|^2\times\\
    & (x_A'=x_A) \times (x_B'=x_B)
   &\text{linearity}\br
\eqv
  & \left|  
      \left(\sum z \cdot 
            (-1)^{(x_A)_z+(x_B)_z} \sqz{\times} 
            H^{\tensor k} \ketb{z} \tensor
            H^{\tensor k} \ketb{z} \right) / \sqrt{2^k} \;
      (y_Ay_B)'\right|^2\times\\
    & (x_A'=x_A) \times (x_B'=x_B)
   &\text{apply } H\br
\eqv
  & \left| 
      \left(\sum z \cdot \right.\right.  
      \begin{aligned}[t]
        & (-1)^{(x_A)_z+(x_B)_z} \sqz{\times}\\
        &(\sum u \cdot (-1)^{z \cdot u}\times\ketb{u}/\sqrt{2^n}) \;\tensor\\
        &(\sum v \cdot (-1)^{z \cdot v}\times\ketb{v}/\sqrt{2^n}) \left.\left.\right)
          / \sqrt{2^k} \;
          (y_Ay_B)'\right|^2\times
      \end{aligned}\\
      & (x_A'=x_A) \times (x_B'=x_B)
      &\text{collect terms}\br
\eqv
  &\left|\sum u,v,z \cdot (-1)^{(x_A)_z + (x_B)_z + u \cdot z + v \cdot z}
    \sqz{\times} \ketb{uv}/{\sqrt{2^k}}^3 \;  (y_Ay_B)'\right|^2\;\times\\
  & (x_A'=x_A) \times (x_B'=x_B)\br
\eqv
  &\left|\sum u,v,z \cdot (-1)^{(x_A)_z \oplus (x_B)_z \oplus (u \oplus v) \cdot z}
    \sqz{\times} \ketb{uv}/{\sqrt{2^k}}^3 \;  (y_Ay_B)'\right|^2\;\times\\
  & (x_A'=x_A) \times (x_B'=x_B)
\end{alignat*}

To demonstrate that $S$ is winning, we need to show $S \lrn P \leq W
\lrn 1$. Let us perform some preliminary calculations first:
\begin{align*}
  & S \times P_0'
  &\text{expand}\br
\eqv
  &\left|\sum u,v,z \cdot (-1)^{(x_A)_z \oplus (x_B)_z \oplus (u \oplus v) \cdot z}
    \sqz{\times} \ketb{uv}/{\sqrt{2^k}}^3 \;  (y_Ay_B)'\right|^2\;\times\\
  & (x_A'=x_A) \times (x_B'=x_B)\times (x_A'=x_B')
  &\text{math}\br
\eqv
  &\left|\sum u,v,z \cdot (-1)^{(u \oplus v) \cdot z}
    \sqz{\times} \ketb{uv}/{\sqrt{2^k}}^3 \;  (y_Ay_B)'\right|^2\;\times\\
  & (x_A'= x_B'= x_A = x_B) 
  &\text{sum}\br
\eqv
  &\left|\sum z \cdot \ketb{zz}/{\sqrt{2^k}} \; (y_Ay_B)'\right|^2 \times
   (x_A'= x_B'= x_A = x_B) 
 &\text{application}\br
\eqv
 & (x_A'= x_B'= x_A = x_B) \times (y_A'=y_B')/2^k\\
\eqv
 & P_0 \times (y_A'=y_B')/2^k \times (x_A'= x_A) \times (x_B'= x_B)
\end{align*}

Similarly, analysing the amplitudes in the second case, we get:
\begin{align*}
  & S \times P_1'
  &\text{expand}\br
\eqv
  & \left|\sum u,v,z \cdot (-1)^{(x_A)_z \xor (x_B)_z \xor (u \xor v) \cdot z}
    \times \ketb{uv} /{\sqrt{2^k}}^3 \; (y_Ay_B)'\right|^2\times\\
  & (x_A'=x_A) \times (x_B'=x_B) \times
    (\sum i \cdot ((x_A')_i =(x_B')_i) \eqv 2^{k-1})
  &\text{split sum}\br
\eqv
  & \left|\left(\sum u \neq v,z \cdot 
      (-1)^{(x_A)_z \xor (x_B)_z \xor (u \xor v) \cdot z}
      \times \ketb{uv} /{\sqrt{2^k}}^3 \right.\right. + \\
  & \hphantom{|(}\left.\left.\sum u,z \cdot 
      (-1)^{(x_A)_z \xor (x_B)_z }
      \times \ketb{uu} /{\sqrt{2^k}}^3 \right)\; (y_Ay_B)'\right|^2\times\\
  & (x_A'=x_A) \times (x_B'=x_B) \times
    (\sum i \cdot ((x_A')_i =(x_B')_i) \eqv 2^{k-1})
  &\text{second sum}\br
\eqv
  & \left|\left(\sum u \neq v,z \cdot 
      (-1)^{(x_A)_z \xor (x_B)_z \xor (u \xor v) \cdot z}
      \times \ketb{uv} /{\sqrt{2^k}}^3 \right.\right. + \\
  & \hphantom{|(}\left.\left.\sum u \cdot 
      ((-1)\times 2^{k-1} + 1 \times 2^{k-1})
      \times \ketb{uu} /{\sqrt{2^k}}^3 \right)\; (y_Ay_B)'\right|^2\times\\
  & (x_A'=x_A) \times (x_B'=x_B) \times
    (\sum i \cdot ((x_A')_i =(x_B')_i) \eqv 2^{k-1})
  &\text{math}\br
\eqv
  & \left|\sum u \neq v,z \cdot 
      (-1)^{(x_A)_z \xor (x_B)_z \xor (u \xor v) \cdot z}
      \times \ketb{uv} /{\sqrt{2^k}}^3 \; (y_Ay_B)'\right|^2\times\\
  & (x_A'=x_A) \times (x_B'=x_B) \times
    (\sum i \cdot ((x_A)_i =(x_B)_i) \eqv 2^{k-1})
  &\text{application}\br
\eqv
  & P_1 \times (y_A' \neq y_B')\times /(2^k\times(2^k-1))
    \times (x_A'=x_A) \times (x_B'=x_B)
\end{align*}

Normalising the winning condition:
\begin{align*}
  & W \lrn 1
  &\text{def of }\lrn\br
\eqv
  & W / \sum \sigma' \cdot W
  &\text{expand}\br
\eqv
  & (P_0 \times (y_A'=y_B') + P_1 \times (y_A' \neq y_B')) /\\
  &\sum y_A',y_B \cdot P_0 \times (y_A'=y_B') + P_1 \times (y_A' \neq y_B')
  &\text{sum}\br
\eqv
  & (P_0 \times (y_A'=y_B') + P_1 \times (y_A' \neq y_B')) /\\
  & (P_0 \times 2^k + P_1 \times 2^k \times (2^k-1))
  &\text{math}\br
\eqv
  & P_0 \times (y_A'=y_B')/2^k + P_1 \times (y_A'\neq y_B')/(2^k \times (2^k-1))
\end{align*}

Finally, the strategy is winning since:
\begin{align*}
  &S \lrn P
  &\text{def of }\lrn\\
\eqv
  &S \times (P_0 + P_1) / (S \seq (P_0 + P_1))
  &\text{expand}\\
\eqv
  &S \times P_0 + S \times P_1 / 
  \sum \sigma'' \cdot (S'' \times (P_0'' + P_1''))
  &\text{above proofs}\\
\eqv
 & P_0 \times (y_A'=y_B')/2^k \times (x_A'= x_A) \times (x_B'= x_B) +\\
 & P_1 \times (y_A' \neq y_B')\times /(2^k\times(2^k-1))
   \times (x_A'=x_A) \times (x_B'=x_B)\\
\eqv
 & (W \lrn 1) \times (x_A'= x_A) \times (x_B'= x_B)\\
\leqv
 & W \lrn 1
\end{align*}

The Deutch-Jozsa game is an example of two-player games. We now turn
our attention to multi-player pseudo-telepathy games.

%%%%%%%%%%%%%%%%%%%%%%%%%%%%%%%%%%%%%%%%%%%%%%%%%%%%%%%%%%%%%%%%%%%%%%%%

\subsection{Mermin's game}
\label{sec:mermin}

In Mermin's game~\cite{mermin90} there are three players.  Each player
$i$ receives a bit $x_i$ as input and outputs a bit $y_i$. The promise
is that the sum of the inputs is even. The players win the game if the
parity of the sum of the outputs is equal to the parity of half the
sum of the inputs.

We formalise the game as follows: $D_i = R_i = \{0,1\}$, for
$i:0,1,2$. The promise is $P \eqv (x_0 \xor x_1 \xor x_2) = 0$. The
winning condition is $W \eqv (y_0' \xor y_1' \xor y_2') = (x_0 + x_1 +
x_2)/2$.

We implement the following quantum strategy. The players share an
entangled state $\psi = \ket{000}/{\sqrt 1} + \ket{111}/\sqrt 2$.
After receiving the input, each player applies the operation $U$
defined by $ U\ket{0} = \ket{0} \text{ and } U\ket{1} =
\sqrt{-1}\times\ket{1}$ to her qubit if the input is $1$. The player
then applies a Hadamard transform. The qubit is measured in the
computational basis and the result of the measurement is the output.

The program is:
\begin{align*}
  &S \eqv \psi \get  \ket{000}/\sqrt{2} + \ket{111}/\sqrt{2}\seq 
          \;  S_0 \;  ||_\psi \;  S_1 \;  ||_\psi \;  S_2\\
  &S_i \eqv
    \textbf{if } x_i=1 \textbf{ then } \psi_i \get  U \psi_i
    \textbf{ else } ok\seq \; 
    \psi_i \get  H \psi_i\seq \; 
    \textbf{measure } \psi_i \;  y_i 
\end{align*}
where $i:0,1,2$.

To prove the solution is correct, we begin by analysing the resulting
distributions of the state variables. As before, we sum out the final
quantum state, since it does not appear in the winning condition.
\begin{align*} 
  &\sum \psi' \cdot S\br
\eqv
  &\sum \psi' \cdot 
  \begin{aligned}[t]
    &\psi \get  \ket{000}/\sqrt{2} + \ket{111}/\sqrt{2}\seq \\
    &||_\psi \;  i:0,1,2 \cdot
    \begin{aligned}[t]
      & \textbf{if } x_i=1 \textbf{ then } \psi_i \get  U \psi_i
      \textbf{ else } ok\seq \\
      & \psi_i \get  H \psi_i\seq \; \textbf{measure } \psi_i \;  y_i
    \end{aligned}
  \end{aligned}
  &\begin{aligned}[t]
    &\\ &\\ &\text{conditional}    
  \end{aligned}\br
\eqv
  &\sum \psi' \cdot 
  \begin{aligned}[t]
    &\psi \get  \ket{000}/\sqrt{2} + \ket{111}/\sqrt{2}\seq \\
    &||_\psi \;  i:0,1,2 \cdot
    \psi_i \get  U^{x_i} \psi_i\seq \psi_i \get  H \psi_i\seq
    \textbf{measure } \psi_i \;  y_i
  \end{aligned}
  &\begin{aligned}[t]
    &\\ &\text{substitute}
  \end{aligned}\br
\eqv
  &\sum \psi' \cdot 
  \begin{aligned}[t]
    &\psi \get  \ket{000}/\sqrt{2} + \ket{111}/\sqrt{2}\seq \\
    &||_\psi \;  i:0,1,2 \cdot
    \textbf{measure } H(U^{x_i}\psi_i) \;  y_i
  \end{aligned}
  &\begin{aligned}[t]
    &\\ &\text{compose}
  \end{aligned}\br
\eqv
  &\sum \psi' \cdot 
  \begin{aligned}[t]
    &\psi \get  \ket{000}/\sqrt{2} + \ket{111}/\sqrt{2}\seq \\
    &\textbf{measure }
    H^{\tensor3}(U^{x_0}\tensor U^{x_1}\tensor U^{x_2}\psi)
    \;  y_0y_1y_2
  \end{aligned}
  &\begin{aligned}[t]
    &&\text{substitute}\\
    &&\text{and measure}
  \end{aligned}\br
\eqv
  &\left|H^{\tensor3}(U^{x_0}\tensor U^{x_0}\tensor U^{x_0}
    (\ket{000}/\sqrt{2}+\ket{111}/\sqrt{2})) \;  (y_0y_1y_3)' \right|^2
    \; \times \\
  &  ((x_0x_1x_2)'=x_0x_1x_2)
  &\text{apply }U\br
\eqv
  &\left|H^{\tensor3}
   (\ket{000}/\sqrt{2}+(\sqrt{-1})^{x_0+x_1+x_2}\times\ket{111}/\sqrt{2})
   \;  (y_0y_1y_3)' \right|^2 \times\\
  &  ((x_0x_1x_2)'=x_0x_1x_2)
\end{align*}

To demonstrate that the strategy $S$ is winning, we need to show $S
\lrn P \leqv W \lrn 1$. We begin with some preliminary calculations:
\begin{alignat*}{2}
 & S \times P' &\text{expand}\br
\eqv
  &\left|H^{\tensor3}
   (\ket{000}/\sqrt{2}+(\sqrt{-1})^{x_0+x_1+x_2}\times\ket{111}/\sqrt{2})
   \;  (y_0y_1y_3)' \right|^2 \times\\
  & ((x_0x_1x_2)'=x_0x_1x_2)\times
    (x_0' \xor x_1' \xor x_2' = 0)
  &\text{split into cases}\br
\eqv
  &\left|H^{\tensor3}
   (\ket{000}/\sqrt{2}+\ket{111}/\sqrt{2})
   \;  (y_0y_1y_3)' \right|^2 \times\\
  & ((x_0x_1x_2)'=x_0x_1x_2)\times
    (x_0' + x_1' + x_2' = 0) \;+\\
  &\left|H^{\tensor3}
   (\ket{000}/\sqrt{2}-\ket{111}/\sqrt{2})
   \;  (y_0y_1y_3)' \right|^2 \times\\
  & ((x_0x_1x_2)'=x_0x_1x_2)\times
    (x_0' + x_1' + x_2' = 2)
  &\text{apply }H\br
\eqv
  &\left|
    (\ket{000}+\ket{011} + \ket{101} + \ket{110})/2
    \;  (y_0y_1y_3)' \right|^2 \times\\
  & ((x_0x_1x_2)'=x_0x_1x_2)\times
    (x_0' + x_1' + x_2' = 0) \;+\\
  &\left|
    (\ket{001}+\ket{010} + \ket{100} + \ket{111})/2
    \;  (y_0y_1y_3)' \right|^2 \times\\
  & ((x_0x_1x_2)'=x_0x_1x_2)\times
    (x_0' + x_1' + x_2' = 2)
  &\text{application}\br
\eqv
  &(y_0' \xor y_1' \xor y_2' = (x_0'+x_1'+x_2')/2)
   \times((x_0x_1x_2)'=x_0x_1x_2)\times P/4
\end{alignat*}

Normalising the winning condition:
\begin{align*}
  &W \lrn 1
  &\text{def of }\lrn\br
\eqv
  &W / \sum \sigma' \cdot W
  &\text{expand}\br
\eqv
  &((y_0' \xor y_1' \xor y_2' = (x_0+x_1+x_2)/2)/\\
  &\sum y_0',y_1',y_2' \cdot y_0' \xor y_1' \xor y_2' = (x_0+x_1+x_2)/2
  &\text{sum}\br
\eqv
  &((y_0' \xor y_1' \xor y_2' = (x_0+x_1+x_2)/2)/4\times P
\end{align*}

Finally, the strategy $S$ is winning, since: 
\begin{align*}
  & S \lrn P 
  &\text{def of }\lrn\br
\eqv
  & S \times P' / (S \seq P)
  &\text{above proofs}\br
\eqv
  &(y_0' \xor y_1' \xor y_2' = (x_0'+x_1'+x_2')/2)
   \times((x_0x_1x_2)'=x_0x_1x_2)\times P/4/\\
   & \sum \sigma'' \cdot
  \begin{aligned}[t]
    &|H^{\tensor3}(\ket{000}/\sqrt{2} +
    (\sqrt{-1})^{x_0+x_1+x_2}\times\ket{111}/\sqrt{2} (y_0y_1y_2)''|^2
    \times\\
    &((x_0x_1x_2)''=x_0x_1x_2)\times(x_0''\xor x_1'' \xor x_2''=0)
  \end{aligned}
  &\text{math}\br
\eqv
  &(y_0' \xor y_1' \xor y_2' = (x_0'+x_1'+x_2')/2)
   \times((x_0x_1x_2)'=x_0x_1x_2)/4\times P
  &\text{def of }W\br
\eqv 
  & (W!1) \times ((x_0x_1x_2)'=x_0x_1x_2)\\
\leqv 
  &(W!1)
\end{align*}

%%%%%%%%%%%%%%%%%%%%%%%%%%%%%%%%%%%%%%%%%%%%%%%%%%%%%%%%%%%%%%%%%%%%%%%%

\subsection{Parity Games}
\label{sec:parity}

In parity games~\cite{brassard03,brassard05,buhrman03} there are at
least three players.  Each player $i$ is given a number $\alpha_i :
0,..2^l$, or, equivalently, an $l$-bit binary string. The promise is
that $\sum i:0,..n \cdot \alpha_i$ is divisible by $2^l$.  Each player
outputs a single bit $\beta_i$. The winning condition is that the sum
of the outputs is half the sum of the inputs $\mod 2$:
\begin{align*}
  &P \eqv (\sum i:0,..n \cdot \alpha_i) \text{ mod } 2^l = 0\\
  &W \eqv (\sum i:0,..n \cdot \beta_i') \text{ mod } 2 = 
          (\sum \alpha_i/2^l) \text{ mod } 2
\end{align*}

Consider the following strategy. The players share an entangled state
$\psi = (\ket{0}^{\tensor n} + \ket{1}^{\tensor n})/\sqrt 2$. Each
player $i$ executes the following program:
\begin{align*}
  & \psi_i \get  U_i \psi_i\seq  \psi_i \get  H \psi_i\seq
    \textbf{measure } \;  \psi_i \;  \beta_i 
\end{align*}
where the operator $U_i$ is defined by 
\begin{equation*}
  U_i\ket{0}=\ket{0} \text{ and } 
  U_i\ket{1} = e^{\pi\times\sqrt{-1}\times\alpha_i/2^{l}}\times \ket{1}
\end{equation*}
and $H$ is the Hadamard transform.

Again, we can prove that $S \lrn P \leq W \lrn 1$, where $S$ refers to
the parallel execution of the above program after the initialisation
of the shared entangled state. As before, we sum out the final quantum
state, since it does not appear in the winning condition:
\begin{align*}
  &\sum \psi' \cdot S\\
  \eqv
  &\sum \psi' \cdot 
  \begin{aligned}[t]
    &\psi \get (\ket{0}^{\tensor n}+\ket{1}^{\tensor n})/\sqrt{2} \seq\\
    &||_i\; \psi_i \get  U_i \psi_i\seq  \psi_i \get  H \psi_i\seq  
    \textbf{measure } \;  \psi_i \;  \beta_i 
  \end{aligned}
  &\begin{aligned}[t]
    &\\&\text{substitute}
  \end{aligned}\br
\eqv
  &\sum \psi' \cdot 
  \begin{aligned}[t]
    &\psi \get (\ket{0}^{\tensor n}+\ket{1}^{\tensor n})/\sqrt{2} \seq\\
    &||_i\; \textbf{measure } \;  H(U_i \psi_i) \;  \beta_i 
  \end{aligned}
  &\begin{aligned}[t]
    &\\ &\text{compose}
  \end{aligned}\br
  \eqv
  &\sum \psi' \cdot 
  \begin{aligned}[t]
    &\psi \get (\ket{0}^{\tensor n}+\ket{1}^{\tensor n})/\sqrt{2} \seq\\
    &\textbf{measure } \;  H^{\tensor n}(U_0\tensor \hdots \tensor U_n \psi) 
    \;  \beta_{0,..n}
  \end{aligned} 
  &\begin{aligned}[t]
    &&\text{substitute}\\ &&\text{and measure}
  \end{aligned}\br
  \eqv
  &\left| H^{\tensor n}(U_0\tensor \hdots \tensor U_n 
    (\ket{0}^{\tensor n}+\ket{1}^{\tensor n})/\sqrt{2}) 
    \;  \beta_{0,..n}' \right|^2 \times\\
  &(\alpha_{0,..n}'=\alpha_{0,..n})
  &\text{apply }U\br
  \eqv
  &\left|H^{\tensor n}
   (\ket{0}^{\tensor n}+
    e^{\pi\times\sqrt{-1}\times\sum i \cdot \alpha_i/2^l}
    \times\ket{1}^{\tensor n})/\sqrt{2} 
    \;  \beta_{0,..n}'\right|^2 \times\\
    &(\alpha_{0,..n}'=\alpha_{0,..n}) 
\end{align*}

Beginning with some preliminary calculations:
\begin{align*}
 & S \times P' &\text{expand}\br
\eqv
  &\left|H^{\tensor n}
   (\ket{0}^{\tensor n}+
    e^{\pi\times\sqrt{-1}\times\sum i \cdot \alpha_i/2^l}
    \times\ket{1}^{\tensor n})/\sqrt{2} 
    \;  \beta_{0,..n}'\right|^2 \times\\
  &(\alpha_{0,..n}'=\alpha_{0,..n}) \times
  ((\sum i:0,..n \cdot \alpha_i') \text{ mod } 2^l = 0)
  &\text{split cases}\br
\eqv
  &((\sum i:0,..n \cdot \alpha_i' /2^l) \text{ mod } 2 = 0)\;\times
   (\alpha_{0,..n}'=\alpha_{0,..n}) \times\\
  &\left|H^{\tensor n}
   (\ket{0}^{\tensor n}+\ket{1}^{\tensor n})/\sqrt{2} 
    \;  \beta_{0,..n}'\right|^2 +\\
  &((\sum i:0,..n \cdot \alpha_i' /2^l) \text{ mod } 2 = 1)\;\times
   (\alpha_{0,..n}'=\alpha_{0,..n}) \times\\
  &\left|H^{\tensor n}
   (\ket{0}^{\tensor n}-\ket{1}^{\tensor n})/\sqrt{2} 
    \;  \beta_{0,..n}'\right|^2 
  &\text{apply }H\br
\eqv
  &((\sum i:0,..n \cdot \alpha_i' /2^l) \text{ mod } 2 = 0)\;\times
   (\alpha_{0,..n}'=\alpha_{0,..n}) \times\\
  &\left|\left(\sum x \cdot (p_x=0)\times\ketb{x}/\sqrt{2^{n-1}}\right)\;
    \beta_{0,..n}'\right|^2 +\\
  &((\sum i:0,..n \cdot \alpha_i' /2^l) \text{ mod } 2 = 1)\;\times
   (\alpha_{0,..n}'=\alpha_{0,..n}) \times\\
  &\left|\left(\sum x \cdot (p_x=1)\times\ketb{x}/\sqrt{2^{n-1}}\right)\;
    \;  \beta_{0,..n}'\right|^2 
  &\text{application} \br
\eqv
  &\left((\sum i:0,..n \cdot \beta_i') \text{ mod } 2 =
         (\sum i:0,..n \cdot \alpha_i /2^l) \text{ mod } 2 \right)
        \times P \;\times \\
  &(\alpha_{0,..n}'=\alpha_{0,..n})/2^{n-1}
\end{align*}
where $p_x$ is defined by
\begin{equation*}
  \text{if } \textbf{x}=x_0x_1\hdots x_{n-1}
  \text{ then } p_x = (\sum i:0,..n \cdot x_i) \text{ mod } 2
\end{equation*}

Finally, the strategy $S$ is winning, since:
\begin{align*}
  & S \lrn P 
  &\text{def }\lrn\br
\eqv
  &S \times P ' / (S \seq P) 
  &\text{above proof}\br
\eqv
  &\left((\sum i:0,..n \cdot \beta_i') \text{ mod } 2 =
         (\sum i:0,..n \cdot \alpha_i /2^l) \text{ mod } 2 \right)
       \times P\;\times\\
  &(\alpha_{0,..n}'=\alpha_{0,..n})/2^{n-1}/\\
  &\sum \sigma'' \cdot 
  \begin{aligned}[t]
    &\left((\sum i:0,..n \cdot \alpha_i'' /2^l) \text{ mod } 2 = 0\right)\;\times
    (\alpha_{0,..n}''=\alpha_{0,..n}) \times\\
    &\left|\left(\sum x \cdot (p_x=0)\times\ketb{x}/\sqrt{2^{n-1}}\right)\;
    \beta_{0,..n}''\right|^2 +\\
    &\left((\sum i:0,..n \cdot \alpha_i'' /2^l)\text{ mod } 2 = 1\right)\times
    (\alpha_{0,..n}''=\alpha_{0,..n}) \times\\
    &\left|\left(\sum x \cdot (p_x=1)\times\ketb{x}/\sqrt{2^{n-1}}\right)\;
    \;  \beta_{0,..n}''\right|^2
  \end{aligned}
  &\begin{aligned}[t]
    &\\&\\&\\&\\&\text{sum}
  \end{aligned}\br
\eqv
  &\left((\sum i:0,..n \cdot \beta_i')\text{ mod }2 = 
         (\sum i:0,..n \cdot \alpha_i /2^l)\text{ mod }2 \right)\;\times\\
  &(\alpha_{0,..n}'=\alpha_{0,..n})/2^{n-1}\times P
  &\text{def of }W\br
\eqv
  &(W \lrn 1) \times (\alpha_{0,..n}'=\alpha_{0,..n})\br
\leqv 
 &W \lrn 1
\end{align*}

Note that if $n=3$ and $l=1$, the parity game is a Mermin game.

%%%%%%%%%%%%%%%%%%% end Quantum Non-locality %%%%%%%%%%%%%%%%%%%%%%%%%

%%%%%%%%%%%%%%%%%%%%%%%%%%%%%%%%%%%%%%%%%%%%%%%%%%%%%%%%%%%%%%%%%%%%%%
% Conclusion
%%%%%%%%%%%%%%%%%%%%%%%%%%%%%%%%%%%%%%%%%%%%%%%%%%%%%%%%%%%%%%%%%%%%%%

\section {Conclusion and Future Work}
\label{sec:conclusion}

We have presented a formal framework for specifying, implementing,
and analysing quantum pseudo-telepathy games. 

Current research focuses on formalising quantum cryptographic
protocols, such as BB84~\cite{bennet84}, in our framework. While the
communication involved in these protocols is amenable to analysis
using tools developed in~\cite{tafliovich09}, the analysis of security
aspects requires additional machinery. Integration of the techniques
developed in this work is one of the more promising directions.

%%%%%%%%%%%%%%%%%%%%%%% end Conclusion %%%%%%%%%%%%%%%%%%%%%%%%%%%%%%%

\bibliographystyle{entcs}
\bibliography{sbmf_entcs}

\appendix

%%%%%%%%%%%%%%%%%%%%%%%%%%%%%%%%%%%%%%%%%%%%%%%%%%%%%%%%%%%%%%%%%%%%%%
% Appendix 
%%%%%%%%%%%%%%%%%%%%%%%%%%%%%%%%%%%%%%%%%%%%%%%%%%%%%%%%%%%%%%%%%%%%%%
\section {Quantum Computation}
\label{sec:qc}

In this section we introduce the basic concepts of quantum mechanics,
as they pertain to the quantum systems that we will consider for
quantum computation. The discussion of the underlying physical
processes, spin-$\frac{1}{2}$-particles, etc.~is not our interest.
We are concerned with the model for quantum computation only.  A
reader not familiar with quantum computing can
consult~\cite{nielsen00} for a comprehensive introduction to the
field.

The \emph{Dirac notation}, invented by Paul Dirac, is often used in
quantum mechanics. In this notation a vector $v$ (a column vector by
convention) is written inside a \emph{ket}: $\ket{v}$.  The dual
vector of $\ket{v}$ is $\bra{v}$, written inside a \emph{bra}. The
inner products are \emph{bra-kets} $\braket{v}{w}$.  For
$n$-dimensional vectors $\ket{u}$ and $\ket{v}$ and $m$-dimensional
vector $\ket{w}$, the value of the inner product $\braket{u}{v}$ is a
scalar and the outer product operator $\ket{v}\bra{w}$ corresponds to
an $m$ by $n$ matrix.  The Dirac notation clearly distinguishes
vectors from operators and scalars, and makes it possible to write
operators directly as combinations of bras and kets.

In quantum mechanics, the vector spaces of interest are the Hilbert
spaces of dimension $2^n$ for some $n \in \mathbb{N}$.  A convenient
orthonormal basis is what is called a \emph{computational basis}, in
which we label $2^n$ basis vectors using binary strings of length $n$
as follows: if $s$ is an $n$-bit string which corresponds to the
number $x_s$, then $\ket{s}$ is a $2^n$-bit (column) vector with $1$
in position $x_s$ and $0$ everywhere else. The tensor product $\ket{i}
\tensor \ket{j}$ can be written simply as $\ket{ij}$.  An arbitrary
vector in a Hilbert space can be written as a weighted sum of the
computational basis vectors.

\begin{description}
\item[Postulate 1 (state space)] Associated to any isolated physical
  system is a Hilbert space, known as the \emph{state space} of the
  system. The system is completely described by its \emph{state
    vector}, which is a unit vector in the system's state space.
\end{description}

\begin{description}
\item [Postulate 2 (evolution)] The evolution of a closed quantum
  system is described by a \emph{unitary transformation}.
\end{description}

\begin{description}
\item [Postulate 3 (measurement)] Quantum measurements are described
  by a collection $\{M_m\}$ of \emph{measurement operators}, which act
  on the state space of the system being measured. The index $m$
  refers to the possible measurement outcomes. If the state of the
  system immediately prior to the measurement is described by a vector
  $\ket{\psi}$, then the probability of obtaining result $m$ is
  $\braket{\psi}{M_m^{\dagger} M_m |\psi}$, in which case the state of
  the system immediately after the measurement is described by the
  vector $\frac{M_m \ket{\psi}}{\sqrt{\braket{\psi}{M_m^{\dagger} M_m
        |\psi}}}$. The measurement operators satisfy the
  \emph{completeness equation} $\sum m \cdot M_m^{\dagger} M_m \eqv I$.
\end{description}

An important special class of measurements is \emph{projective
  measurements}, which are equivalent to general measurements provided
that we also have the ability to perform unitary transformations.

A projective measurement is described by an \emph{observable} $M$,
which is a Hermitian operator on the state space of the system being
measured. This observable has a spectral decomposition $M=\sum m \cdot
\lambda_m \times P_m$, where $P_m$ is the projector onto the
eigenspace of $M$ with eigenvalue $\lambda_m$, which corresponds to
the outcome of the measurement.  The probability of measuring $m$ is
$\braket{\psi}{P_m | \psi}$, in which case immediately after the
measurement the system is found in the state $\frac{P_m
  \ket{\psi}}{\sqrt{\braket{\psi}{P_m | \psi}}}$.

Given an orthonormal basis $\ket{v_m}$, $0 \leq m < 2^n$, measurement
with respect to this basis is the corresponding projective measurement
given by the observable $M = \sum m \cdot \lambda_m \times P_m$, where
the projectors are $P_m = \ket{v_m}\bra{v_m}$.

Measurement with respect to the computational basis is the simplest
and the most commonly used class of measurements. In terms of the
basis $\ket{m}$, $0 \leq m < 2^n$, the projectors are $P_m =
\ket{m}\bra{m}$ and $\braket{\psi}{P_m | \psi} = |\psi_m|^2$. The
state of the system immediately after measuring $m$ is $\ket{m}$.

For example, measuring a single qubit in the state $\alpha \times
\ket{0} + \beta \times \ket {1}$ results in the outcome $0$ with
probability $|\alpha|^2$ and outcome $1$ with probability $|\beta|^2$.
The state of the system immediately after the measurement is $\ket{0}$
or $\ket{1}$, respectively.

Suppose the result of the measurement is ignored and we continue the
computation. In this case the system is said to be in a \emph{mixed
  state}. A mixed state is not the actual physical state of the
system. Rather it describes our knowledge of the state the system is
in.  In the above example, the mixed state is expressed by the
equation $\ket{\psi} = |\alpha|^2 \times \{\ket{0}\} + |\beta|^2
\times \{\ket{1}\}$. The equation is meant to say that $\ket{\psi}$ is
$\ket{0}$ with probability $|\alpha|^2$ and it is $\ket{1}$ with
probability $|\beta|^2$. An application of operation $U$ to the mixed
state results in another mixed state, $U(|\alpha|^2 \times \{\ket{0}\}
+ |\beta|^2 \times \{\ket{1}\}) = |\alpha|^2 \times \{U\ket{0}\} +
|\beta|^2 \times \{U\ket{1}\}$.

\begin{description}
\item [Postulate 4 (composite systems)] The state space of a composite
  physical system is the tensor product of the state spaces of the
  component systems. If we have systems numbered $0$ up to and
  excluding $n$, and each system $i$, $0 \leq i < n$, is prepared in
  the state $\ket{\psi_i}$, then the joint state of the composite
  system is $\ket{\psi_0} \tensor \ket{\psi_1} \tensor \ldots \tensor
  \ket{\psi_{n-1}}$.
\end{description}

While we can always describe a composite system given descriptions of
the component systems, the reverse is not true. Indeed, given a state
vector that describes a composite system, it may not be possible to
factor it to obtain the state vectors of the component systems. A
well-known example is the state $\ket{\psi} = \ket{00}/\sqrt{2} +
\ket{11}/\sqrt{2}$. Such a state is called an \emph{entangled} state.
%%%%%%%%%%%%%%%%%%%%%%%%% end Appendix %%%%%%%%%%%%%%%%%%%%%%%%%%%%%%%

\end{document}

%% file: defs.tex
%%%%%%%%%%%%%%%%%%%%%%%%%%%%%%%%%%%%%%%%%%%%%%%%%%%%%%%%%%%%%%%%%%%%%%
%% definitions
%%%%%%%%%%%%%%%%%%%%%%%%%%%%%%%%%%%%%%%%%%%%%%%%%%%%%%%%%%%%%%%%%%%%%%

%%%%%%%%%%%%%%%%%%%%%%%%%%%%%%%%%%%%%%%%%%%%%%%%%%%%%%%%%%%%%%%%%%%%%%
%% definitions
%%%%%%%%%%%%%%%%%%%%%%%%%%%%%%%%%%%%%%%%%%%%%%%%%%%%%%%%%%%%%%%%%%%%%%

% long equals sign
\newcommand{\eqv}{\boldsymbol{=}\!\boldsymbol{=} \;}

% long left implication arrow
\newcommand{\lar}{\boldsymbol{\Longleftarrow} \;}

% right implication arrow
\newcommand{\impl}{\Rightarrow}

% long right implication arrow
\newcommand{\rar}{\boldsymbol{\Longrightarrow} \;}

% bold less-than-or-equals
\newcommand{\leqv}{\;\;\boldsymbol{\leq}\;\;}
% bold greater-than-or-equals
\newcommand{\geqv}{\;\;\boldsymbol{\geq}\;\;}

% assignment
\newcommand{\get}{\!:=\!}

% sequential composition
\newcommand{\seq}{\; ; \;}

% parallel composition
\newcommand{\parr}{\; || \;}

% measure
\newcommand{\measure}[3]{\textbf{measure}_{#3} \; #1 \; #2 }

% chan, qchan
\newcommand{\chan}{\textbf{chan }}
\newcommand{\qchan}{\textbf{qchan }}

% channel output 
\newcommand{\out}{\,!\,}

% channel output 
\newcommand{\lrn}{\,!\,}

% decrease space around operator
\newcommand{\sqz}[1]{\!#1\!}

% var
\newcommand{\var}{\textbf{var }}

% if-then-else
\newcommand{\ifthenelse}[3]{\textbf{if } #1 \textbf{ then } #2
  \textbf{ else } #3}
% if-then-
\newcommand{\tifthen}[2]{\textbf{if } #1 \textbf{ then } #2}
% -else
\newcommand{\telse}{\textbf{else }}

%% a new ``break'' command: to allow page breaks in align and similar
% environments 
\newcommand{\br}{\displaybreak[0] \\}

% tensor product
\newcommand{\tensor}{\otimes}

% xor operator
\newcommand{\xor}{\oplus}

% a nice list-like environment
\newcommand{\entrylabel}[1]{%
                 \mbox{\textbf{#1:}}\hfil}
\newenvironment{entry}%
  {\begin{list}{}{%
    \renewcommand{\makelabel}{\entrylabel}%
    \setlength{\labelwidth}{35pt}%
    \setlength{\leftmargin}%
                 {\labelwidth+\labelsep}}}%
  {\end{list}}

% ket, bra, and braket
\newcommand{\ket}[1]{|#1\rangle}
\newcommand{\bra}[1]{\langle#1|}
\newcommand{\braket}[2]{\langle #1 | #2 \rangle}

% bold ket, bra, and braket
\newcommand{\ketb}[1]{|\textbf{#1}\rangle}
\newcommand{\brab}[1]{\langle\textbf{#1}|}
\newcommand{\braketb}[2]{\langle \textbf{#1} | \textbf{#2} \rangle}

% a bad shortcut
\newcommand{\ip}{i\!\!+\!\!1}

% how much space from left margin to equations
% default set by fleqn option is 2.5pt
%\mathindent=15pt

\newcommand{\subsec}[1]{\bigskip\noindent\textbf{#1    }}
%%%%%%%%%%%%% end definitions %%%%%%%%%%%%%%%%%%%%%%%%%%%%%%%%%%%%%%%%